\documentclass[preprint,showpacs,amsmath,amssymb,aip,cha]{revtex4-1}

\usepackage{graphicx}
\usepackage{dcolumn}
\usepackage{bm}

\begin{document}

\title{High-frequency and high-field optically detected magnetic resonance of nitrogen-vacancy centers in diamond} 

\author{Viktor Stepanov}
\affiliation{Department of Chemistry, University of Southern California, Los Angeles CA 90089, USA}

\author{Franklin H. Cho}
\affiliation{Department of Physics, University of Southern California, Los Angeles CA 90089, USA}

\author{Chathuranga Abeywardana}
\affiliation{Department of Chemistry, University of Southern California, Los Angeles CA 90089, USA}

\author{Susumu Takahashi}
\email{susumu.takahashi@usc.edu}
\affiliation{Department of Chemistry, University of Southern California, Los Angeles CA 90089, USA}
\affiliation{Department of Physics, University of Southern California, Los Angeles CA 90089, USA}

\date{\today}

\begin{abstract}
We present the development of an optically detected magnetic resonance (ODMR) system, which enables us to perform the ODMR measurements of a single defect in solids at high frequencies and high magnetic fields. Using the high-frequency and high-field ODMR system, we demonstrate 115 GHz continuous-wave and pulsed ODMR measurements of a single nitrogen-vacancy (NV) center in a diamond crystal at the magnetic field of 4.2 Tesla as well as investigation of field dependence ($0-8$ Tesla) of the longitudinal relaxation time ($T_1$) of NV centers in nanodiamonds.
\end{abstract}


\maketitle 

A nitrogen-vacancy (NV) center is a paramagnetic color center in diamond with unique electronic, spin, and optical properties
including its stable fluorescent (FL) signals~\cite{Gruber97}
and long decoherence time.~\cite{Kennedy03, Gaebel06, Childress06, Balasubramanian09, Takahashi08}
Moreover, it is possible to initialize the spin states of NV centers by applying optical excitation and to readout the states by measuring the FL intensity.~\cite{Jelezko04}
Electron spin resonance (ESR) of a single NV center is observable by measuring changes of the FL intensity, a magnetic resonance technique known as optically detected magnetic resonance (ODMR) spectroscopy.~\cite{Gruber97}
In addition, NV centers are extremely sensitive to their surrounding electron~\cite{Gaebel06, Hanson06, Hanson08} and nuclear spins.~\cite{Childress06, Balasubramanian09}
Therefore a NV center in diamond is a promising candidate for investigation of fundamental sciences
\cite{Gruber97, Jelezko04, Epstein05, Gaebel06, Childress06, Hanson08, Takahashi08, Balasubramanian09, Wang13}
and for applications to quantum information processing~\cite{Dutt07, Fuchs09, Neumann10, Bernien13}
and magnetic sensing.~\cite{Degen08, Balasubramanian08, Maze08, Taylor08, Maletinsky12, Grinolds13, Mamin13sci, Staudacher13, Ohashi13, muller14}

Key motivations for the NV magnetic sensing include the detection of a single spins and improvement of sensitivity of ESR and nuclear magnetic resonance (NMR) spectroscopies to the level of a single spin.
For example, detection of a single or a small ensemble of electron and nuclear spins surrounding a NV center has been demonstrated using the double electron-electron resonance (DEER) and electron-nuclear double resonance (ENDOR) spectroscopy of a single NV center at low magnetic fields.~\cite{deLange12, Mamin12, Laraoui12, Mamin13sci, Sushkov14}
In addition, spin relaxometry based on the longitudinal relaxation time ($T_1$) measurement of a single NV center has been employed to detect several electron spins.~\cite{Steinert13, Kaufmann13}

In this article, we present the development of a high-frequency (HF) ODMR system to investigate NV centers in diamond.
Similar to NMR spectroscopy, the spectral resolution of ODMR and NV-based magnetic resonance (MR) techniques including DEER and ENDOR is significantly improved at HF,
thus highly advantageous in distinguishing target spins from other species ({\it e.g.}, impurities existing in diamond) for NV-based MR spectroscopy.
In addition, HF MR spectroscopy can produce extremely high spin polarization at low temperatures,
which improves the signal-to-noise ratio of NV-based MR measurements on ensembles of target electron spins
and increases spin coherence of target spins significantly.~\cite{Takahashi08, Takahashi09prl, Takahashi11, Edwards12, wang11}
The HF ODMR employs a confocal FL imaging system, a HF excitation component and a 12.1 Tesla superconducting magnet.
The HF excitation component is a part of the HF ESR spectrometer operating in the frequency range of 107-120 GHz and 215-240 GHz,~\cite{Cho14} therefore the system is capable of performing {\it in-situ} HF ODMR and HF ESR experiments.
We also present HF ODMR measurements of a single NV center in diamond.
First, a single NV center is identified by FL imaging, autocorrelation, and field-dependent FL measurements.
Then, we perform continuous-wave (cw) ODMR at the external magnetic field of 4.2 Tesla and microwave frequency of 115 GHz.
We also demonstrate the coherent manipulation of a single NV center spin at 4.2 Tesla by performing Rabi oscillations, pulsed ODMR, and spin echo measurements.
Finally, we perform measurements of the longitudinal relaxation time ($T_1$) of NV centers in nanodiamonds (NDs) using the HF ODMR system.
The experiment shows that $T_1$ in NDs is shorter than $T_1$ in bulk diamonds, and is nearly field-independent in the range of $0-8$ Tesla.
The observation agrees with $T_1$ process due to paramagnetic spins surrounding NV centers in NDs.

\begin{figure}
\includegraphics[width=100 mm]{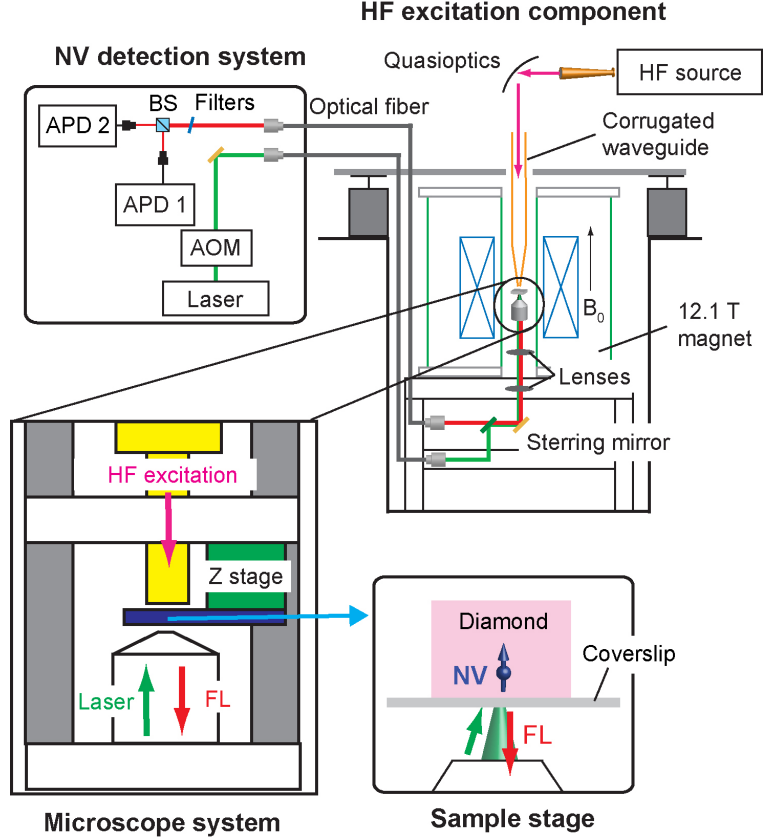}
\caption{
Overview of the HF ODMR system. The HF source in the HF excitation component is tunable continuously in the range of 107-120 GHz and 215-240 GHz.
HF microwaves are guided by quasioptics and a corrugated waveguide.
The NV detection system consists of a 532 nm cw diode-pumped solid state laser, an AOM, fiber couplers, optical filters, a beam splitter (BS), and APDs.
The excitation laser is applied to NV centers through a microscope objective located at the center of the 12.1 Tesla superconducting magnet and the FL signals of NV centers are collected by the same objective. The FL signals are filtered by optical filters in the NV detection system. For autocorrelation measurements, the FL signals are split into two and detected by two separate APDs. The microscope system consists of a microscope objective, a z-translation stage, and the corrugated waveguide. The sample stage is supported by the z-translation stage.
}
\end{figure}
Figure 1 shows an overview of the HF ODMR system consisting of a NV detection system, a HF excitation component, a microscope system, and a sample stage.
A cw 532 nm laser in the NV detection system is employed for optical excitation of NV centers.
The cw excitation laser first couples to an acousto-optic modulator (AOM) for pulsed operations, then is transmitted through a single-mode optical fiber to the bottom of a 12.1 Tesla superconducting magnet where the excitation laser couples to the microscope system.
As shown in Fig.~1, the microscope system consists of a microscope objective (Zeiss) for both the optical excitation and collection of NV's FL signals, a z-translation stage (Attocube), and a corrugated waveguide (Thomas-Keating) for the HF microwave excitation.
The HF microwave excitation component is a part of our HF ESR spectrometer.~\cite{Cho14}
The output frequency of the HF microwave source is continuously tunable in the range of 107-120 GHz and 215-240 GHz.
The sample stage is supported by the z-translation stage.
The z-direction of the sample position is adjusted by the z-translation stage, and the xy-direction of the laser excitation volume is controlled by a combination of a sterring mirror and a pair of lenses.
The HF ODMR employs a confocal microscope system to detect FL signals of NV centers.
The FL signals are collected by the same objective, then are transmitted to the detection system by a 50 $\mu$m multi-mode optical fiber cable that enables confocal FL imaging.
Finally the FL signals are filtered by optical filters and detected by avalanche photodiodes (APDs) in the NV detection system.

\begin{figure}
\includegraphics[width=140 mm]{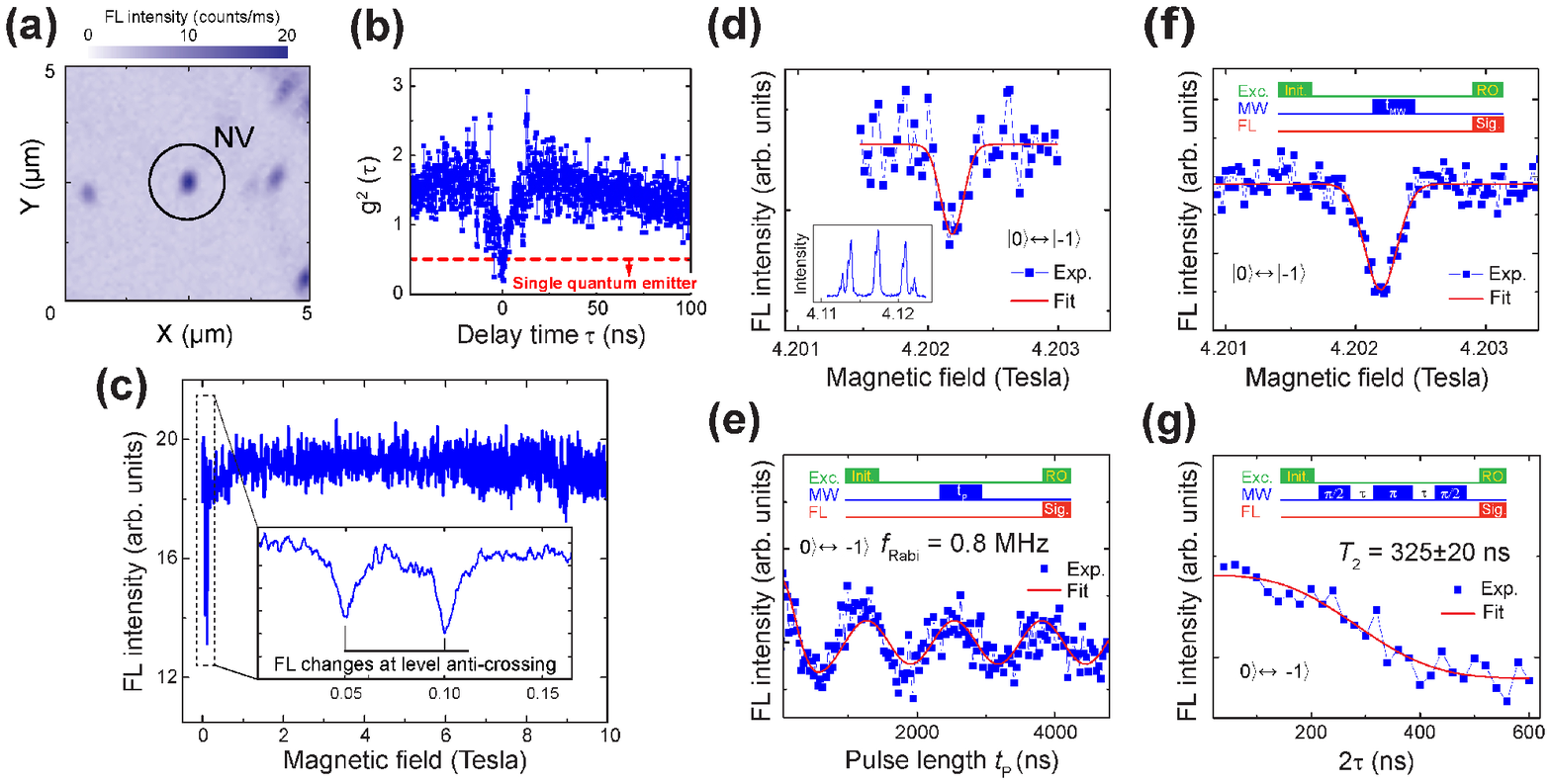}
\caption{
(a) FL intensity image of a type-Ib diamond crystal. The scanning area is 5 $\times$ 5 $\mu$m$^2$. Solid circle indicates a single NV center that was used in the subsequent measurements. (b) Autocorrelation curve observed from the the single NV center. The observation of $g^2(\tau=0)<0.5$ confirms the detection of the single NV center.
(c) Magnetic field dependence of the single NV center FL intensity. The field was applied along the (111) axis of the diamond within 8 degrees. The inset shows reduction of FL intensity at $\sim$0.05 and 0.1 Tesla due to LAC of the excited and ground states of the NV center, respectively.
(d) cw ODMR measurement of the single NV center at 115 GHz. The ODMR signal of the single NV center was observed at 4.2022 Tesla. The solid line indicates a fit to the Gaussian function. The inset shows the {\it in-situ} ESR measurement of single-substitutional nitrogen impurities in the diamond.
(e) Rabi oscillation experiment. The frequency of the observed Rabi oscillations was 0.8 MHz. The inset shows the applied pulse sequence consisting of the initialization (Init.) and readout (RO) pulses by the 532 nm laser (Exc.), microwave pulse (MW) of length $t_P$, and FL signals (Sig.). Init.=4 $\mu$s and RO=Sig.=300 ns were used in the measurement. $t_P$ was varied.
(f) Pulsed ODMR as a function of magnetic field. The solid line indicates a fit to the Gaussian function. The inset shows the pulse sequence. Init.=4 $\mu$s, RO=Sig.=300 ns, and $t_{MW}$=500 ns were used in the measurement.
(g) The spin echo measurement to determine the spin decoherence time ($T_2$) of the single NV center. The solid line indicates a fit to $exp(-(2\tau/T_2)^3)$.~\cite{Wang13} The inset shows the pulse sequence. Init.=4 $\mu$s, RO=Sig.=300 ns, $\pi/2$=250 ns, and $\pi$=600 ns were used in the measurement. $\tau$ was varied.
The pulse sequence was repeated on the order of $10^6$ times to obtain a single point in all measurements.
}
\end{figure}
Using the HF ODMR system, we performed FL measurements on a single crystal of type-Ib diamond (1.5 $\times$ 1.5 $\times$ 1.1 mm$^3$, Sumitomo Electric Industries). First, as shown in Fig. 2a, FL imaging was carried out to map FL signals from the diamond crystal.
After choosing a well-isolated FL peak (Fig.~2a), we performed anti-bunching measurement and observed the autocorrelation signal which verifies that the FL signal was due to a single quantum emitter (Fig.~2b).
Then, the FL intensity was monitored continuously as we applied the external magnetic field from 0 to 10 Tesla.
As shown in the inset of Fig.~2c, we observed two dips of the FL intensity at $\sim$0.05 and 0.1 Tesla, originating from the level anti-crossing (LAC) in NV's optically excited and ground states, respectively.~\cite{Epstein05}
Therefore the observation of the LAC and autocorrelation signal confirmed the detection of a single NV center.
In addition, as shown in Fig.~2c, we found that the FL intensity is stable in high magnetic fields up to 10 Tesla.

Next, we demonstrated ODMR measurements of the single NV center at the microwave frequency of 115 GHz.
First, we performed cw ODMR spectroscopy of the single NV center with applications of cw microwave and laser excitations.
As shown in Fig.~2d, the observed FL signals as a function of magnetic field showed a dip of the FL intensity at 4.2022 Tesla, which corresponds to the $m_S=0 \leftrightarrow -1$ transition of the NV center.
The $g$-factor of the NV center was estimated to be 2.0027-2.0041 by taking into account for the strength of magnetic field at the resonance and uncertainty of the orientation of magnetic field ($<8$ degrees).
The calibration of magnetic field was done by {\it in-situ} ensemble ESR measurement of single-substitutional nitrogen impurities (see the inset of Fig.~2d).~\cite{Takahashi08}
Then, we carried out pulsed experiments at 115 GHz and 4.2022 Tesla.
In the pulsed experiments, the NV center was first prepared in the $m_S=0$ spin sublevel with the application of the laser initialization pulse, then the microwave excitation pulse sequence was applied.
The final state of the NV center was determined by measuring its FL intensity with the application of the laser readout pulse.
In all pulsed measurements, the FL intensity was normalized by the reference signal, which is the FL intensity measured without the microwave pulse sequence in order to cancel noises associated with laser intensity fluctuations and thermal/mechanical instability of the setup.
In addition, each data point was obtained by repeating the pulse sequence and averaging FL signals on the order of $10^6$ times.
As shown in Fig.~2e, Rabi oscillations of the same single NV center were observed by varying the duration of the single microwave pulse ($t_{P}$).
Figure~2f shows pulsed ODMR signals of the single NV center as a function of magnetic field.
The observed full-width at half-maximum was 0.29 mT (8 MHz) which is typical for NV centers in type-Ib diamond crystals.~\cite{Abeywardana14}
Next, we measured the spin decoherence time ($T_2$) of the single NV center at 4.2022 Tesla.
As shown in the inset of Fig.~2g, the applied microwave sequence consists of the spin echo sequence and an additional $\pi/2$ pulse that converts the resultant coherence of the NV center into the $m_S=0$ state.~\cite{Jelezko04}
By fitting the observed FL decay to $exp((-2\tau/T_2)^3)$,~\cite{Wang13} $T_2$ of the single NV center was determined as 325$\pm$20 ns (see Fig.~2g).

\begin{figure}
\includegraphics[width=80 mm]{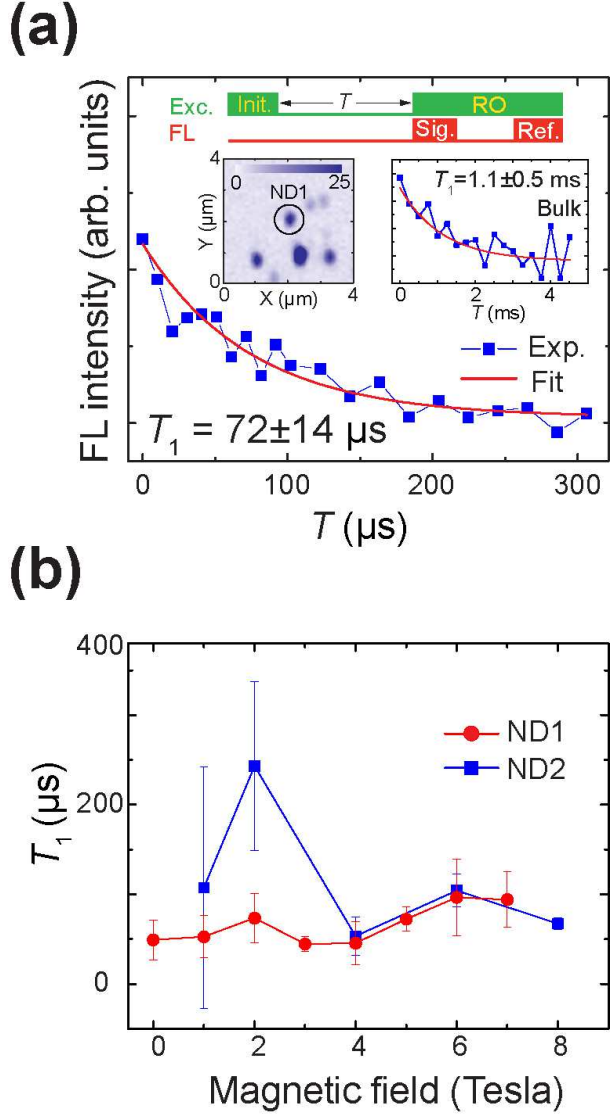}
\caption{
$T_1$ measurements of NV centers in NDs.
(a) $T_1$ measurement of NV centers in a ND (ND1). The solid line indicates a fit to the single exponential decay function, $exp(-T/T_1)$. A FL image of ND1 is shown in the inset. The inset also shows the pulse sequence consisting of the initialization (Init.) and readout (RO) pulses by the 532 nm laser (Exc.), and FL signals (Sig.).
$T$ was varied. Init.=4 $\mu$s, RO=3 $\mu$s, and Sig.=300 ns were used. The reference (Ref.) of 300 ns was measured 2 $\mu$s after the signal.
The pulse sequence was repeated on the order of $10^6$ times to obtain a single point in all measurements.
In addition, the inset shows $T_1$ measurement of a single NV center in the type-Ib bulk diamond obtained using the same pulse sequence above ($T_1 = 1.1 \pm 0.5$ ms).
(b) $T_1$ as a function of magnetic field for two NDs (ND1 and ND2).
}
\end{figure}
Finally, using our HF ODMR system, we studied the field dependence of $T_1$ of NV centers in NDs.
The samples we studied here were NV-enhanced type-Ib NDs with the average diameter of 35 nm (Academia Sinica).~\cite{Chang08}
The NDs were sparsely placed on a coverslip, and identified by performing FL images ({\it e.g.}, the inset of Fig.~3a for ND1) and by observing the LAC signals from the field-dependent FL measurements.
Then $T_1$ relaxation measurements were performed as shown in Fig.~3a.
The pulse sequence employed for $T_1$ measurements is shown in the inset of Fig.~3a.
The NV centers in NDs were first prepared in the $m_S=0$ spin sublevel with the application of the laser initialization pulse, then a dark time interval ($T$) was applied with no laser excitation.
The final state of the NV centers was determined by measuring the FL intensity of the NV centers with the application of the laser readout pulse.
The FL signal with the laser excitation was used as a reference signal for the normalization.
In order to detect the $T_1$ relaxations from the $m_S=0$ state to the thermal equilibrium state, the normalized FL signal was measured as a function of $T$.
As shown in Fig.~3a, an exponential decay of the FL intensity was clearly observed.
From a single exponential fit of the FL signal, we obtained $T_1=72\pm$14 $\mu$s.
The measured $T_1$ is 10$-$100 times shorter than typical $T_1$ in bulk diamonds (see the inset of Fig.~3a).~\cite{Takahashi08, Neumann08, Steinert13}
The shorter $T_1$ is often observed in NV centers in NDs and near the diamond surface, and
it has been reported that transverse magnetic field fluctuations from paramagnetic impurities located on or nearby the diamond surface provide an additional contribution to $T_1$ of NV centers.~\cite{Steinert13, Kaufmann13, Tetienne13,Rosskopf14, Myers14}
Because the intensity of the fluctuations depends on the difference of the Larmor frequencies between the NV centers and the impurities,~\cite{Steinert13}
the contribution to $T_1$ will be nearly independent of the strength of magnetic field for the impurities observed in NDs having $g$-values$\sim$2.\cite{Laraoui12, Casabianca11}
Moreover, the $T_1$ measurements were performed at several magnetic fields for two different NDs.
As shown in Fig.~3b, we observed similar values of $T_1$ for both NDs in the range of $0-8$ Tesla, which supports the $T_1$ process due to paramagnetic spins.~\cite{Steinert13, Kaufmann13, Tetienne13, Myers14}

%
%
In summary, we presented the development of the HF ODMR system, which enables us to perform the ODMR measurements of a single defect in solids at high frequencies and high magnetic fields. Using the HF ODMR system, we demonstrated cw ODMR, Rabi oscillation, pulsed ODMR, and spin echo measurements of a single NV center in a type-Ib diamond crystal at the microwave frequency of 115 GHz and magnetic field of 4.2 Tesla, and studied $T_1$ of NV centers in NDs as a function of magnetic field.

%
%
We would like to thank Steven Retzloff, Jeffrey Hesler, Stephen Jones (Virginia Diodes Inc.), and Don Wiggins (USC machine shop) for technical assistance on the development of the HF ODMR system. This work was supported by the Searle scholars program (S.T.).

\end{document}